\documentclass[
    aps,
    prb,
    twocolumn,
    superscriptaddress,
    floatfix,
    longbibliography,
    showkeys,
    10pt
]{revtex4-2}
\usepackage[english]{babel}
\babelprovide[hyphenrules=english]{en}
\usepackage{amsmath,amssymb,mathtools,bm}
\usepackage{graphicx}
\usepackage{physics}
\usepackage{siunitx}
\usepackage{csquotes}
\usepackage[table]{xcolor}
\usepackage{soul}

\sethlcolor{yellow!45}
\soulregister\cite7
\soulregister\ref7
\soulregister\textsc7
\soulregister\texttt7
\soulregister\textquote7
\soulregister\AA0

\usepackage[hidelinks]{hyperref}
\graphicspath{{./IMG/}}

\usepackage{hyperref} 
\begin{document}
\title{First-Principles Investigation of Ammonia Adsorption and N--H Activation on W--B Nanoclusters}

\author{Anastasiia V. Iosimovska (corresponding author)}
\email{anastasiia.iosimovska@skoltech.ru}
\affiliation{Skolkovo Institute of Science and Technology, Bolshoy Boulevard 30, bld. 1, Moscow 121205, Russian Federation}

\author{Mikhail M. Lukanov} 
\affiliation{Skolkovo Institute of Science and Technology, Bolshoy Boulevard 30, bld. 1, Moscow 121205, Russian Federation}

\author{Christian Tantardini (corresponding author)}
\email{christiantantardini@ymail.com}
\affiliation{Center for Integrative Petroleum Research, King Fahd University of Petroleum and Minerals, Dhahran 31261, Saudi Arabia}

\author{Alexander S. Novikov} 
\affiliation{Institute of Natural Sciences, Immanuel Kant Baltic Federal University, Alexandra Nevskogo St., 14, Kaliningrad, 236041, Russian Federation}

\author{Viktor S. Baidyshev}
\affiliation{Skolkovo Institute of Science and Technology, Bolshoy Boulevard 30, bld. 1, Moscow 121205, Russian Federation}

\author{Alexander G. Kvashnin (corresponding author)}
\email{A.Kvashnin@skoltech.ru}
\affiliation{Skolkovo Institute of Science and Technology, Bolshoy Boulevard 30, bld. 1, Moscow 121205, Russian Federation}

\begin{abstract}
Tungsten-boride nanoclusters are nanoscale systems of potential catalytic interest, but their structural stability and reactivity remain poorly understood. We combine evolutionary structure prediction with density functional theory to investigate W$_m$B$_n$ nanoclusters containing up to 43 atoms. The calculated stability descriptors reveal a strongly non-monotonic landscape, with several compositions locally favoured relative to neighbouring stoichiometries within the explored isolated-cluster model. Molecular NH$_3$ adsorption was examined at W-top and accessible B-top sites, with adsorption energies ranging from $-1.78$ to $-0.54$~eV on W sites and from $-1.54$ to $+0.62$~eV on B sites. H adsorption at W-top and B-top sites spans $-1.47$ to $+0.67$~eV, indicating pronounced composition- and site-dependence. CI-NEB calculations show that the first N--H bond-cleavage barrier is strongly pathway-dependent. These results characterize the initial activation of NH$_3$ on W--B nanoclusters and provide an atomistic basis for further investigation of the complete ammonia-decomposition mechanism.
\end{abstract}

\keywords{ammonia decomposition, hydrogen storage,
tungsten boride nanoclusters, density functional theory, USPEX}
 
\maketitle

\section{Introduction}

Hydrogen is increasingly viewed as both an energy carrier and an industrial feedstock for decarbonization, particularly in sectors where direct electrification is difficult.\cite{staffell2019role,yue2021hydrogen} Its high gravimetric energy density of about 142~MJ~kg$^{-1}$, together with the possibility of electricity generation in fuel cells with water as the main reaction product, makes it attractive for transport, stationary power, grid balancing, oil refining, ammonia and methanol synthesis, and low-carbon steel production.\cite{sarker2023prospect,yue2021hydrogen}

Despite this promise, present hydrogen production remains dominated by fossil-fuel routes. According to the IEA Global Hydrogen Review 2025, global hydrogen production reached almost 100~Mt H$_2$ in 2024, while low-emissions hydrogen accounted for less than 1\% of total supply and unabated fossil-fuel pathways generated roughly 980~Mt CO$_2$.\cite{IEA2025HydrogenReview} The direct use of molecular hydrogen also faces practical limitations associated with its low volumetric density, the need for compression at 350--700~bar or liquefaction at $-253^{\circ}$C, leakage and flammability risks, and compatibility problems with metallic infrastructure, including embrittlement and corrosion.\cite{zhang2023hydrogen,mekonnin2025hydrogen,calabrese2024hydrogen,MULKY2024129710}

These limitations have motivated strong interest in chemical hydrogen-rich carriers. Among them, ammonia is a particularly attractive candidate because it contains 17.6--17.8~wt\% hydrogen, has a volumetric hydrogen density higher than that of compressed hydrogen, avoids the cryogenic storage requirements of liquid hydrogen, and benefits from an already established production, storage, and transport infrastructure.\cite{portarapillo2025ammonia,chatterjee2021limitations} At the same time, ammonia must be treated not only as a carrier but also as a reactive pollutant. Agricultural systems are already the dominant source of atmospheric NH$_3$, mainly through livestock housing, manure storage and spreading, and urea- or ammonium-based fertilization.\cite{EEA2025AmmoniaAgriculture,Ma2021AmmoniaMitigation} Once emitted, NH$_3$ contributes to secondary ammonium aerosols and reduced-nitrogen deposition, with consequences for fine particulate matter formation, eutrophication, soil acidification, biodiversity loss, and inefficient nitrogen use.\cite{Sutton2008AmmoniaEnvironment,Aneja2009AgricultureAirClimate,Erisman2007ReducedNitrogen} This does not weaken the case for ammonia as a hydrogen carrier, but it makes containment, NH$_3$ slip suppression, recovery as ammonium salts, or controlled conversion into N$_2$ and H$_2$ essential technological requirements.\cite{Moore2018AirScrubber}

This work in general focus on catalytic hydrogen generation from ammonia decomposition,
\[
\mathrm{2NH_3 \rightarrow 3H_2 + N_2},
\]
but more specific we pay attention on the first step of ammonia decomposition. Although the reaction is thermodynamically favoured at elevated temperature and low pressure, practical hydrogen release is limited by reaction kinetics, residence time, and residual NH$_3$ slip.\cite{lucentini2021review,schuth2012ammonia} Catalyst development therefore aims to increase NH$_3$ conversion and H$_2$ productivity at reduced operating temperature, under realistic gas hourly space velocities and with sufficient stability against deactivation.\cite{lamb2019ammonia} Reported activation energies for representative supported catalysts span approximately 19--142~kJ~mol$^{-1}$, depending on the active phase, support, particle size, and reaction conditions.\cite{han2025catalyst}

At the microscopic level, ammonia decomposition is commonly rationalized through the competition between NH$_3$ activation and the removal of surface nitrogen species. Following the Sabatier principle, N-containing intermediates must bind strongly enough to enable N--H bond cleavage, but not so strongly that the surface becomes poisoned by N adatoms.\cite{bell2016h2} This balance gives rise to volcano-type activity trends that relate catalytic performance to nitrogen binding energy.\cite{ganley2004priori,boisen2005optimal} Ru-based catalysts lie close to this optimum and remain benchmark low-temperature systems. For example, Ru/CNT and Cs-promoted Ru/CNT catalysts have been reported to reach H$_2$ production rates of 6353~$\mathrm{mol}~\mathrm{H}_2~\mathrm{mol}_{\mathrm{Ru}}^{-1}~\mathrm{h}^{-1}$ at 430~$^{\circ}$C and 7870~$\mathrm{mol}~\mathrm{H}_2~\mathrm{mol}_{\mathrm{Ru}}^{-1}~\mathrm{h}^{-1}$ at 370~$^{\circ}$C, respectively.\cite{bell2016h2} However, the cost and scarcity of Ru, and in some cases the use of alkali promoters such as Cs, limit their large-scale practicality and motivate the search for more accessible catalyst chemistries.\cite{su2023review}

Non-noble Ni-, Fe-, and Co-based catalysts, bimetallic systems, carbides, nitrides, and related transition-metal compounds have therefore been widely investigated, but many of these systems still suffer from insufficient low-temperature activity, structure sensitivity, stability limitations, or non-optimal binding of reaction intermediates.\cite{lucentini2021review,bell2016h2,han2025catalyst,ghoreishian2025recent} Recent reviews identify transition-metal carbides and nitrides, including WC$_x$ and WN$_x$ phases, as relevant catalyst families for ammonia decomposition.\cite{ghoreishian2025recent}

Within W-containing materials, tungsten carbide and tungstated zirconia have been directly studied for dilute NH$_3$ decomposition.\cite{pansare2007ammonia} Tungsten carbide was reported to achieve complete decomposition of 4000~ppm NH$_3$ at 550~$^{\circ}$C after H$_2$/CO pretreatment, but its behaviour was strongly affected by pretreatment and surface reconstruction. A subsequent study showed that tungsten carbide and tungstated zirconia were active for dilute NH$_3$ decomposition and, under selected conditions, outperformed a commercial Fe-based catalyst.\cite{pansare2008ammonia} However, in syngas at 600~$^{\circ}$C, tungsten carbide showed no conversion, whereas tungstated zirconia and Amomax-10 gave approximately 20\% and 10\% conversion, respectively.\cite{pansare2008ammonia} These results show that W-based compounds can participate in NH$_3$ decomposition, but their activity depends strongly on phase identity, surface state, pretreatment, and reaction atmosphere. Tungsten carbide systems have therefore not provided a general solution for efficient low-temperature hydrogen release from ammonia.\cite{schuth2012ammonia}

This motivates the exploration of W-based chemistries beyond carbide-, nitride-, and oxide-derived phases. Tungsten borides are a chemically distinct class of candidate materials whose potential for ammonia decomposition remains largely unexplored. More generally, transition-metal borides combine metal--metal, metal--boron, and boron--boron bonding, leading to broad compositional flexibility, tunable electronic structure, high thermal stability, and variable local coordination environments.\cite{pu2021nanostructured,carenco2013nanoscaled} These features are relevant because ammonia decomposition is highly sensitive to local surface structure and to the adsorption strength of NH$_3$, H, and NH$_x$ intermediates.\cite{bell2016h2,boisen2005optimal} In W--B systems, W centres may provide transition-metal $d$ states for NH$_3$ adsorption and N--H bond activation, while W--B bonding can modify the electronic structure and coordination of the active sites and create B-containing environments that influence the stabilization of H- and NH$_x$-containing fragments. At the nanoscale, W--B clusters can additionally expose a high fraction of low-coordinated W and B atoms, edge and corner sites, and composition-dependent local motifs that are less accessible in extended bulk phases.\cite{baletto2005structural}

Although direct studies of W--B nanoclusters for ammonia decomposition are lacking, the broader catalytic potential of higher tungsten borides has already been demonstrated. For example, WB$_{5-x}$ used as a cocatalyst with TiO$_2$ increased photocatalytic H$_2$ production by a factor of 23 relative to pure TiO$_2$.\cite{kurenkova2024photocatalytic} 
At the same time, recent first-principles studies of low-dimensional nanomaterials have shown how atomistic structure, charge localization, dimensionality, and functional response can be directly connected by computational simulations.\cite{chepkasov_structure-driven_2024,chepkasov_core-dictated_2025,Tantardini2025SbIV}
Here, we present a first-principles investigation of small W--B clusters
containing up to 43 atoms, focusing on their structure, stability, adsorption behaviour, and the initial N--H bond cleavage step in
NH$_3$ decomposition.
The calculated activation barriers provide a basis for assessing composition- and site-dependent trends in the initial N--H activation step and for guiding further investigation of W--B systems in ammonia decomposition.

\section{Methods}

\subsection{Global optimization with USPEX}

Low-energy structures of W$_m$B$_n$ nanoclusters were explored using the evolutionary algorithm implemented in \textsc{USPEX} \cite{glass2006uspex,lyakhov2013new}. Structural relaxations during the evolutionary searches were performed with \textsc{VASP}. Searches were carried out in the zero-dimensional variable-composition cluster mode \cite{lepeshkin2018method,lepeshkin2022magic}, which allows different cluster sizes and stoichiometries to be explored within a selected compositional range and related structural motifs to be inherited across neighbouring compositions.

The explored composition space covered W$_{1-20}$B$_{1-25}$ clusters. Because the stability descriptors defined below require neighbouring compositions, they were evaluated for $m \leq 19$ and $n \leq 24$, i.e., for clusters with $m+n \leq 43$. To improve sampling efficiency, this range was divided into $4\times5$ composition windows, such as W$_{1-5}$B$_{1-5}$ and W$_{1-5}$B$_{6-10}$, giving 20 independent search blocks. Periodic boundary conditions were used only as a computational framework for the isolated clusters.

Each variable-composition search was initialized with 400 structures, followed by generations containing 200 structures. Parent selection used a best-fraction parameter of 0.8, and up to 140 low-energy structures were retained in the evolutionary pool. New candidates were generated using heredity, random symmetric generation, softmutation, transmutation, atom addition, and atom removal. Antiseeds were used to preserve structural diversity and reduce repeated sampling of already explored regions of configuration space. The convergence of the searches for the selected compositions is shown in Fig.~S1 of the Supporting Information.

Reduced-accuracy DFT settings were used during the evolutionary stage to accelerate configurational sampling. These calculations employed the projector augmented-wave method,\cite{Blochl1994PAW,Kresse1999PAW} a plane-wave cutoff of 250~eV, spin polarization, first-order Methfessel--Paxton smearing with a width of 0.10~eV, and $\Gamma$-point sampling. Symmetry was switched off. Low-energy structures obtained from USPEX were then re-optimized using the higher-accuracy settings described below.

\subsection{DFT calculations}

All DFT calculations, including structural relaxations during the evolutionary search, final total-energy calculations, structural optimizations, adsorption calculations, and reaction-path calculations, were performed with \textsc{VASP} code.\cite{Kresse1993VASP,Kresse1994VASP,Kresse1996VASP_CMS,Kresse1996VASP_PRB} Spin-polarized density functional theory was used throughout within the projector augmented-wave formalism.\cite{Blochl1994PAW,Kresse1999PAW} Competing spin states were screened for the bare cluster and one representative NH$_3$-adsorbed structure of each of the selected compositions; the relative energies of the screened spin states and the magnetic moments are listed in Tables~S4 and S5 of the Supporting Information, respectively. Exchange--correlation effects were described using the RPBE generalized-gradient approximation,\cite{Hammer1999revPBE} and long-range dispersion interactions were included through Grimme's D3 correction.\cite{Grimme2010D3}

The Kohn--Sham orbitals were expanded in a plane-wave basis with a kinetic-energy cutoff of 480~eV. Electronic occupations were treated using Gaussian smearing with a width of 0.05~eV, and the electronic self-consistency threshold was set to $10^{-5}$~eV. All final structural optimizations were continued until the residual force on each atom was below 0.02~eV~\AA$^{-1}$. The nanoclusters were modelled in periodically repeated supercells with at least 15~\AA\ of vacuum between periodic images. Brillouin-zone sampling was restricted to the $\Gamma$ point, as appropriate for isolated finite clusters in large supercells.

Adsorption sampling was performed at top sites on both elements. For molecular NH$_3$, the analyzed dataset contains 113 initially W-top, N-down configurations and 173 B-top configurations covering the symmetry-inequivalent geometrically accessible W and B atoms of the 21 selected clusters. For atomic H, 130 W-top and 174 B-top initial configurations were calculated. In every calculation, both the adsorbate and all cluster atoms were allowed to relax, so adsorption-induced displacement or reconstruction was not constrained. Configuration-level data are reported in Tables~S1 and S2.

The molecular-ammonia adsorption energy was calculated as
\begin{equation}
E_{\mathrm{ads}}(\mathrm{NH_3})=
E_{\mathrm{cluster+NH_3}}-E_{\mathrm{cluster}}-E_{\mathrm{NH_3}},
\end{equation}
whereas H adsorption was referenced to molecular hydrogen,
\begin{equation}
E_{\mathrm{ads}}(\mathrm{H})=
E_{\mathrm{cluster+H}}-E_{\mathrm{cluster}}-\frac{1}{2}E_{\mathrm{H_2}}.
\end{equation}
Zero-point vibrational and finite-temperature entropic corrections were not included.

Reaction pathways for the first N--H bond cleavage of adsorbed NH$_3$ were calculated using the climbing-image nudged elastic band (CI-NEB) method,\cite{Jonsson1998NEB,Henkelman2000NEB,Henkelman2000CI} as implemented in \textsc{VASP}. Each pathway connected separately optimized initial (IS) and final (FS) states and contained eight intermediate images. Final states were constructed by transferring one H atom from adsorbed NH$_3$ to nearby W or B atoms. The nearest acceptor atom in the IS was considered preferentially, while selected neighbouring sites were additionally examined to sample alternative local environments. The highest-energy climbing-image configuration was treated as an estimate of the transition state and is denoted TS throughout the text.

\section{Results and discussion}

\subsection{Predicted stable structures of W\texorpdfstring{$_m$}{ m}B\texorpdfstring{$_n$}{ n} clusters}

\begin{figure*}[t]
    \centering
    \includegraphics[width=\textwidth]{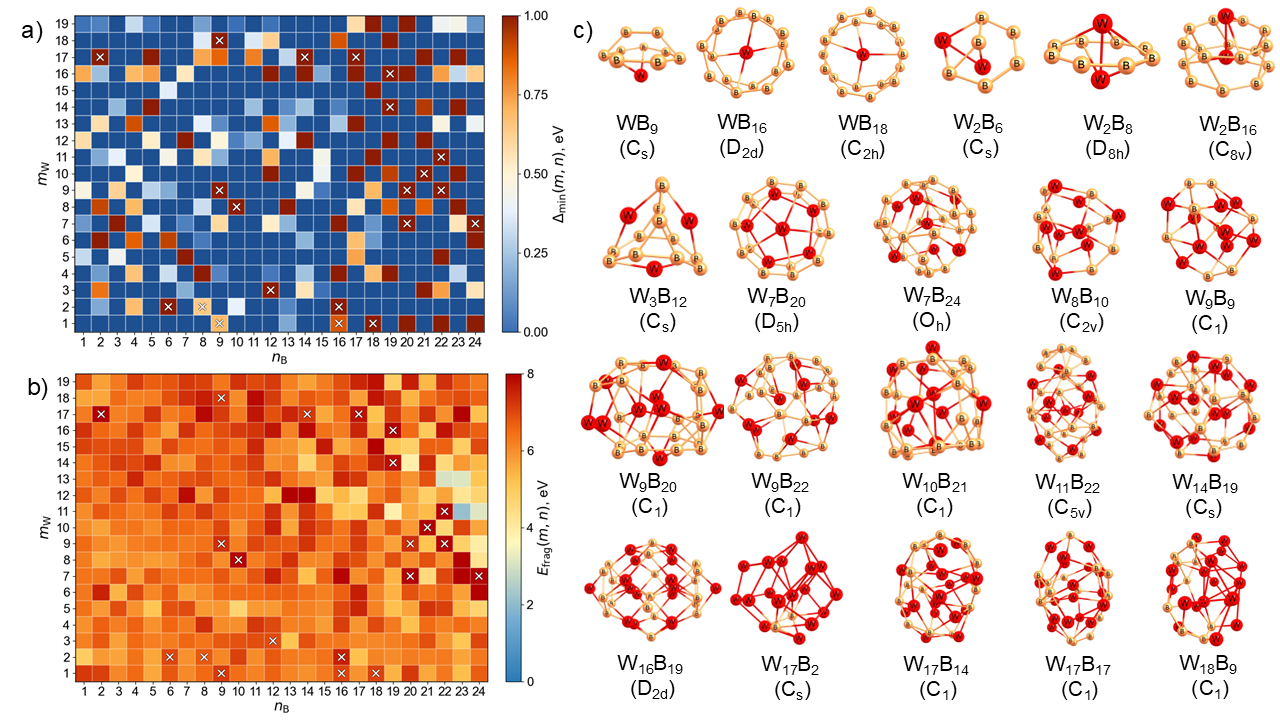}
    \caption{Stability and structural evolution of W$_m$B$_n$ nanoclusters with $m+n \leq 43$.
a) Local compositional stability map expressed by the second-order finite-difference descriptor $\Delta_{\min}(m,n)$.
b) Minimum fragmentation energy map $E_{\mathrm{frag}}(m,n)$.
c) Selected low-energy structures of representative W$_m$B$_n$ clusters together with their compositions and point-group symmetries.}
    \label{fig:clusters}
\end{figure*}

The variable-composition USPEX search yielded low-energy W$_m$B$_n$ clusters across the investigated W$_{1-20}$B$_{1-25}$ composition range. These optimized structures serve as the basis for the stability, adsorption, and reaction-path analyses discussed below.
The resulting stability maps and representative low-energy structures of W$_m$B$_n$ nanoclusters with $m+n \leq 43$ are presented in Fig.~\ref{fig:clusters}.
Figure~\ref{fig:clusters}a shows the map of local compositional stability expressed through the second-order finite-difference descriptor $\Delta_{\min}(m,n)$.
Fig.~\ref{fig:clusters}b presents the minimum fragmentation energy $E_{\mathrm{frag}}(m,n)$, and Fig.~\ref{fig:clusters}c displays selected structures (highlighted by $\times$ in Fig.~\ref{fig:clusters}a) together with their compositions and point-group symmetries.

To quantify the relative energetic stability of the generated W$_m$B$_n$ nanoclusters, within the explored compositional space, we used two complementary descriptors derived from the energies of the lowest-energy structures identified for neighbouring compositions. 
The first one is the local compositional stability \cite{lepeshkin2018method}
\begin{equation}
\Delta_{\min}(m,n)=\min\{\Delta_{\mathrm{W}}(m,n),\Delta_{\mathrm{B}}(m,n)\},
\end{equation}
where
\begin{equation}
\Delta_{\mathrm{W}}(m,n)=E(m+1,n)+E(m-1,n)-2E(m,n),
\end{equation}
\begin{equation}
\Delta_{\mathrm{B}}(m,n)=E(m,n+1)+E(m,n-1)-2E(m,n),
\end{equation}
and $E(m,n)$ is the energy of the lowest-energy W$_m$B$_n$ structure identified in the search.
Positive values of $\Delta_{\min}(m,n)$ identify compositions that are locally favoured relative to neighbouring stoichiometries with respect to atom transfer. In the present work, the term ``magic composition'' refers to compositions showing enhanced local stability according to the $\Delta_{\min}$ descriptor within the explored composition space.

The second descriptor is the minimum fragmentation energy \cite{lepeshkin2018method}
\begin{equation}
E_{\mathrm{frag}}(m,n)=\min_{k,l}\left[E(k,l)+E(m-k,n-l)-E(m,n)\right],
\end{equation}
obtained by considering all non-trivial binary dissociation channels of the type W$_m$B$_n \rightarrow$ W$_k$B$_l$ + W$_{m-k}$B$_{n-l}$. In contrast to $\Delta_{\min}$, which probes local stability in composition space, $E_{\mathrm{frag}}$ characterizes the calculated resistance of an isolated cluster to binary fragmentation into smaller W--B fragments. A larger positive value therefore corresponds to a less favourable fragmentation process within the present isolated-cluster model.

The maps in Fig.~\ref{fig:clusters}a,b reveal a highly non-monotonic relative-stability landscape composed mainly of isolated stability islands rather than a simple trend with cluster size or W:B ratio. The $\Delta_{\min}(m,n)$ map in Fig.~\ref{fig:clusters}a is rather selective, with only a limited set of compositions exhibiting pronounced positive maxima, indicating enhanced local stability relative to neighbouring stoichiometries. These compositions can therefore be regarded as locally stable, or \textquote{magic}, within the explored compositional space. By contrast, the $E_{\mathrm{frag}}(m,n)$ map in Fig.~\ref{fig:clusters}b is overall more uniformly positive, indicating that many clusters are energetically resistant to the considered binary fragmentation channels even when their local compositional stability is only moderate. This difference is useful because a cluster may be resistant to fragmentation but still compete energetically with nearby compositions. Thus, Fig.~\ref{fig:clusters}a identifies compositions that are locally preferred within the explored W--B composition space, while Fig.~\ref{fig:clusters}b characterizes their energetic resistance to binary fragmentation. Together, these two descriptors characterize complementary aspects of the relative stability of the isolated W--B clusters considered here.
The structures shown in Fig.~\ref{fig:clusters}c demonstrate that the low-energy structural motifs identified in the search vary strongly with composition.
B-rich clusters such as WB$_{16}$, WB$_{18}$, W$_2$B$_{16}$, and W$_3$B$_{12}$ are dominated by extended boron frameworks with ring- and cage-like motifs, while tungsten atoms occupy embedded, capping, or bridge-like positions within the boron scaffold. 
At intermediate compositions, for example W$_2$B$_8$, W$_8$B$_{10}$, W$_9$B$_9$, and W$_{11}$B$_{22}$, the structures become distinctly three-dimensional and combine compact W-containing polyhedra with more open boron fragments. 
In the most W-rich clusters, especially W$_{17}$B$_2$ and W$_{18}$B$_9$, the tungsten sublattice forms the main structural backbone, whereas boron acts as a stabilizing dopant occupying surface and bridge positions. 
The analysis of the point groups shows that only selected compositions retain relatively high symmetry, such as WB$_{16}$ (D$_{2d}$), W$_2$B$_8$ (D$_{8h}$), W$_7$B$_{20}$ (D$_{5h}$), W$_7$B$_{24}$ (O$_h$), W$_{11}$B$_{22}$ (C$_{5v}$), and W$_{16}$B$_{19}$ (D$_{2d}$), whereas many other clusters relax to lower-symmetry C$_s$ or C$_1$ structures. 
The predominance of low-symmetry structures indicates substantial structural diversity across the explored composition space, while several compositions retain distinct high-symmetry low-energy motifs.
This structural diversity can be understood in terms of the competition between W--W interactions, W--B bonding, and boron-based multicentre bonding.

\subsection{Adsorption properties}

\begin{figure*} [t]
    \centering
    \includegraphics[width=1\linewidth]{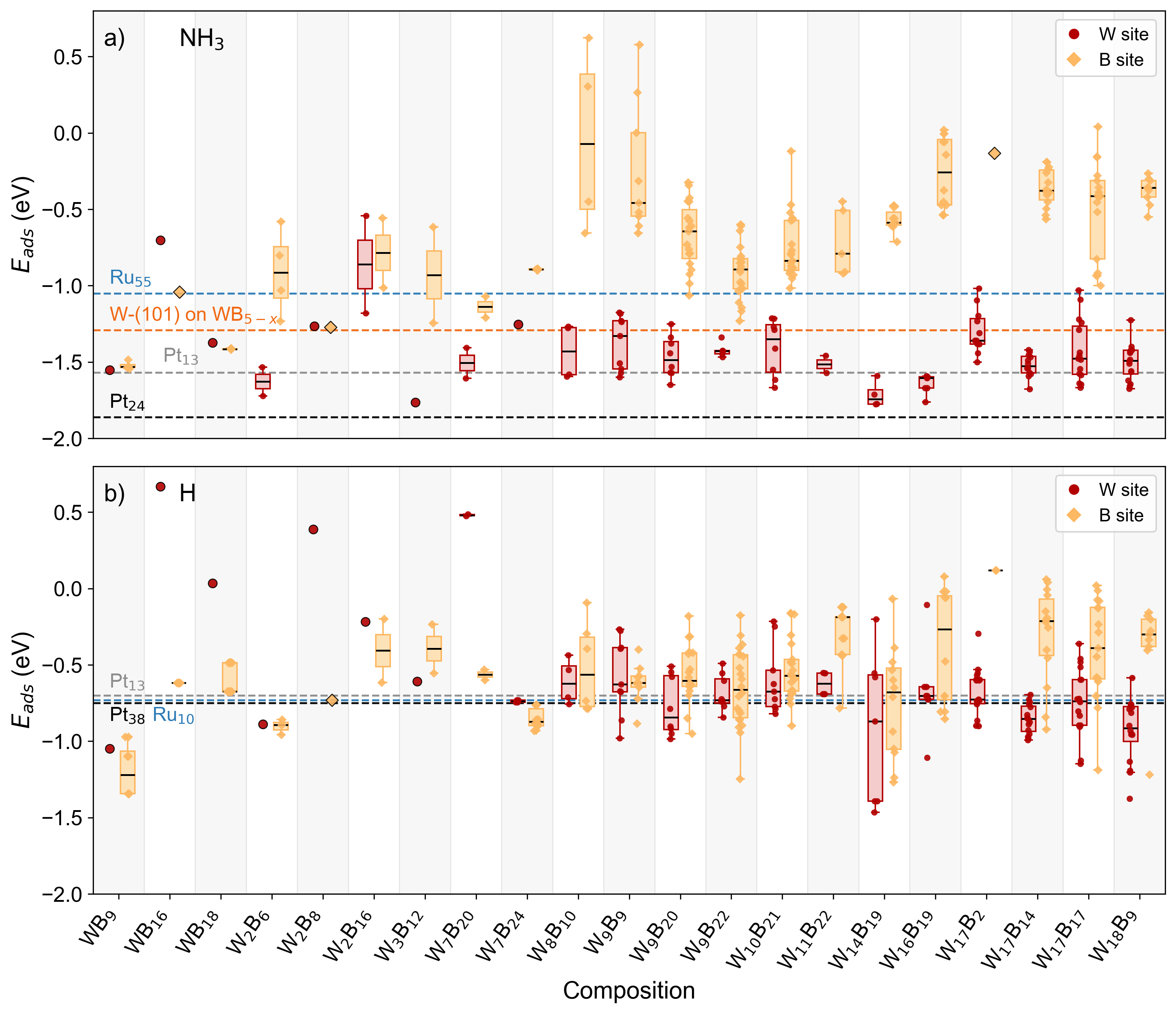}
\caption{Adsorption energies of (a) NH$_3$ and (b) H on W--B nanoclusters. Red circles and orange diamonds denote configurations initialized at W-top and B-top sites, respectively. Box plots show the adsorption-energy distributions for each composition. Horizontal dashed lines indicate literature reference values for NH$_3$ adsorption on Ru$_{55}$ \cite{chattaraj2024adsorption}, W-terminated WB$_{5-x}$(101) \cite{radina2024theoretical}, and Pt$_{13}$ and Pt$_{24}$ clusters \cite{kadioglu2014investigation}, and for H adsorption on a gas-phase Ru$_{10}$ cluster \cite{chen2015hydrogen} and Pt$_{13}$ and Pt$_{38}$ clusters \cite{okamoto2006comparison}. These cross-system reference lines are included for qualitative context because the cluster models and computational protocols differ.}
\label{fig:adsorption}
\end{figure*}

The adsorption analysis was used to characterize molecular NH$_3$ binding and the stabilization of H produced in the first N--H bond-cleavage step. The complete site-resolved values and optimized host--adsorbate distances are given in Tables~S1 and S2, and their composition-resolved distributions are summarized in Fig.~\ref{fig:adsorption}.

All configurations initialized at W-top sites remain N--W coordinated after relaxation. The NH$_3$ adsorption energies range from $-1.78$ to $-0.54$~eV, with a median of $-1.47$~eV. The strongest adsorption is found for W$_{14}$B$_{19}$ ($E_{\mathrm{ads}}\approx-1.78$~eV) and W$_3$B$_{12}$/W$_{16}$B$_{19}$ ($E_{\mathrm{ads}}\approx-1.76$~eV), whereas the weakest value is obtained for W$_2$B$_{16}$ ($E_{\mathrm{ads}}\approx-0.54$~eV). The optimized W--N distances span 2.23--2.53~\AA.

NH$_3$ adsorption was also examined at B-top sites. The relaxed configurations remain N--B coordinated, with adsorption energies ranging from $-1.54$ to $+0.62$~eV and a median of $-0.61$~eV. The optimized B--N distances span 1.60--1.76~\AA. For W$_8$B$_{10}$, for example, the B-top adsorption energies range from approximately $-0.66$ to $+0.62$~eV, whereas the corresponding W-top values lie between approximately $-1.59$ and $-1.27$~eV. W-top sites bind NH$_3$ more strongly, although adsorption at B-top sites is also observed.

The distributions further show that NH$_3$ adsorption depends not only on the host element but also on its local atomic environment. The literature values shown in Fig.~\ref{fig:adsorption}a fall within or close to the range obtained here; however, this comparison remains qualitative because the reference systems differ in size, composition, and computational treatment.

H adsorption, referenced to one half of gas-phase H$_2$, also shows a pronounced site dependence (Fig.~\ref{fig:adsorption}b). W-top configurations span $-1.47$ to $+0.67$~eV, with a median of $-0.73$~eV, whereas the B-top configurations span $-1.34$ to $+0.12$~eV, with a median of $-0.57$~eV. The substantial overlap between the two distributions indicates that H stabilization is determined by both the host element and the local environment. Depending on the site, H binding can vary from strong to weak or unfavorable on both W and B atoms. This variation is directly relevant to the first N--H bond-cleavage step, because the reaction energetics depend not only on the initial NH$_3$ geometry but also on the local site available to accommodate the detached H atom.

\subsection{First N--H bond cleavage in adsorbed NH\texorpdfstring{$_3$}{3}}

After characterizing NH$_3$ and atomic H adsorption, the first elementary step of NH$_3$ decomposition is examined by Nudged Elastic Band (NEB) calculations \cite{Henkelman2000NEB}. 
The considered reaction corresponds to the cleavage of one N--H bond in adsorbed ammonia:
\[
\mathrm{NH_3^*} \rightarrow \mathrm{NH_2^* + H^*},
\]
where the asterisk denotes an adsorbed species. 
The initial states are taken from optimized molecular NH$_3$ adsorption geometries. 
For each selected initial state, one or several dissociated final states are constructed by placing the detached H atom at different nearby adsorption sites, followed by geometry optimization. A single NH$_3$ adsorption geometry can give several distinct NEB pathways, depending on the final position of H and the local W--B environment.

For each pathway, the forward activation barrier is defined as
\begin{equation}
E_{\mathrm{a,fwd}} = E_{\mathrm{TS}} - E_{\mathrm{IS}},
\end{equation}
while the reverse barrier is calculated as
\begin{equation}
E_{\mathrm{a,rev}} = E_{\mathrm{TS}} - E_{\mathrm{FS}}.
\end{equation}
The reaction energy is evaluated as
\begin{equation}
\Delta E = E_{\mathrm{FS}} - E_{\mathrm{IS}}.
\end{equation}
Here, IS corresponds to molecularly adsorbed NH$_3$, TS is the climbing-image configuration, and FS corresponds to co-adsorbed NH$_2$ and H fragments. 
The full set of IS, TS, and FS energies, reaction energies, and activation barriers is reported in Table~S3 of the Supporting Information, together with the transferred H atom and the H-accepting atom in the relaxed FS for each pathway.

\begin{figure*}[t]
\centering
\includegraphics[
    width=0.82\textwidth
]{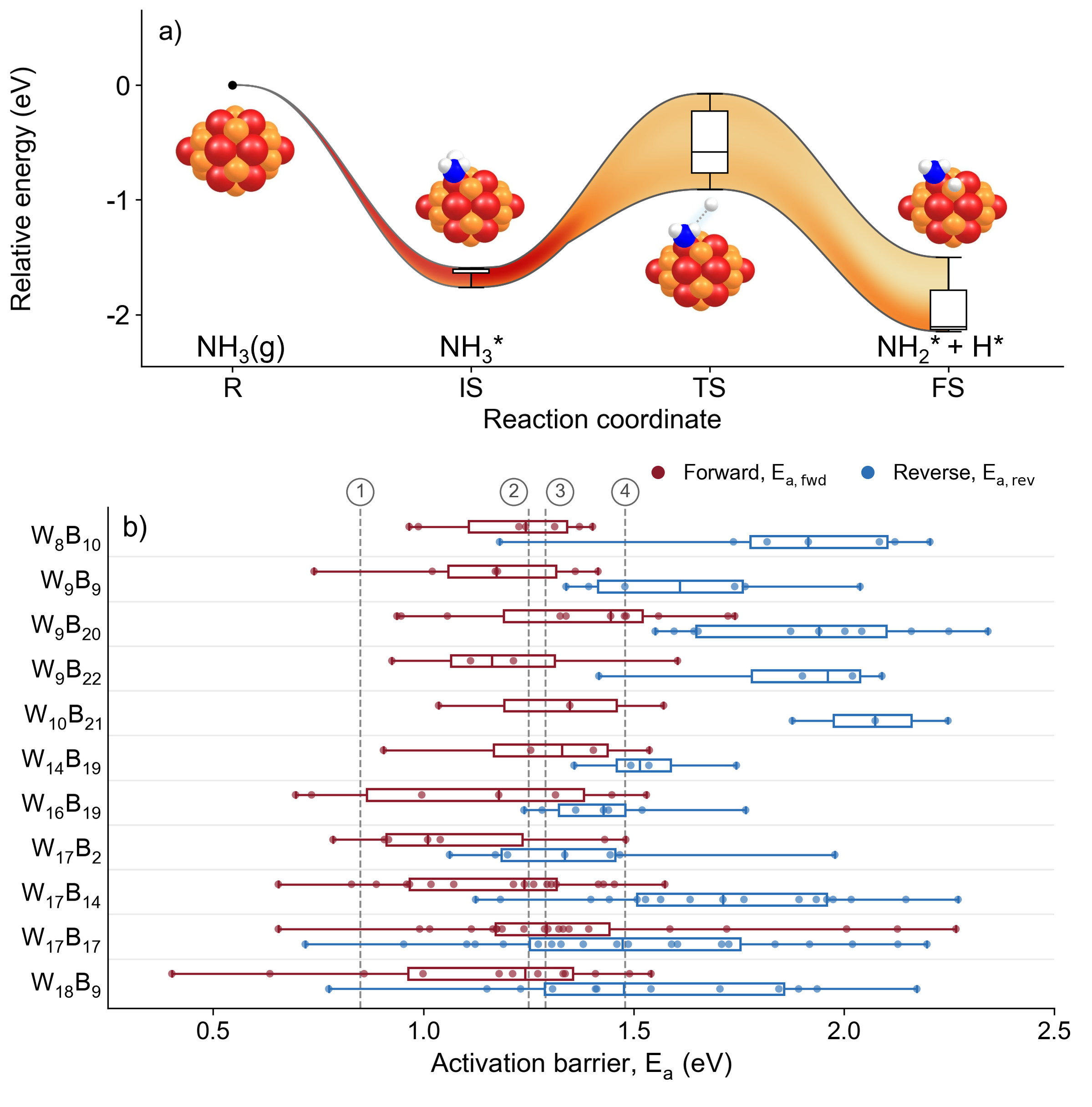}
\caption{Representative first N--H bond cleavage pathway and activation-barrier distributions for ammonia decomposition on W--B nanoclusters. (a) Reaction-energy profile for the W$_{16}$B$_{19}$ cluster. The zero-energy reference R is the separated W$_{16}$B$_{19}$ cluster and gas-phase NH$_3$. IS, TS, and FS correspond to molecularly adsorbed NH$_3^*$, the climbing-image configuration, and the dissociated NH$_2^*$ + H$^*$ state, respectively. The asterisk denotes adsorbed species. (b) Distributions of calculated forward and reverse activation barriers for selected multichannel W--B nanocluster compositions. Red and blue boxplots correspond to $E_{a,\mathrm{fwd}}$ and $E_{a,\mathrm{rev}}$, respectively; individual points represent separate CI-NEB reaction channels. Grey dashed vertical lines indicate literature reference barriers for the first N--H cleavage step: 1 -- Ru$_{13}$/CNT, 0.85~eV, from Zhou et al. \cite{zhou2018first}; 2 -- Pt$_3$/graphene, 1.25~eV, from Cui et al. \cite{cui2019enhanced}; {3 -- bare Ni$_6$, 1.29~eV, from Miao et al. \cite{miao2021graphene};} and 4 -- Fe$_{55}$, 1.48~eV, from Lanzani and Laasonen \cite{lanzani2010nh3}. The literature values are shown for qualitative comparison; the Pt$_3$/graphene value corresponds to a free-energy barrier including zero-point vibrational energy and entropy corrections, whereas the present W--B values are electronic CI-NEB barriers.}
\label{fig:neb_barriers}
\end{figure*}

\begin{table*}[t]
\caption{Sampling statistics and energetics of the CI-NEB pathways for the first N--H bond cleavage on W--B nanoclusters with at least three sampled initial states ($N_{\mathrm{IS}} \geq 3$). $N_{\mathrm{IS}}$ and $N_{\mathrm{path}}$ are the numbers of sampled initial states and pathways; H$_{\mathrm{FS}}$ (B/W) gives the number of pathways in which the atom nearest to the transferred H in the relaxed final state is B or W. Energies are in eV; SD is the sample standard deviation. Data for all individual pathways are listed in Table~S3.}
\label{tab:neb}
\begin{ruledtabular}
\begin{tabular}{lcccccccc}
Cluster & $N_{\mathrm{IS}}$ & $N_{\mathrm{path}}$ & H$_{\mathrm{FS}}$ (B/W) & $\langle E_{\mathrm{a,fwd}}\rangle \pm$ SD & Median $E_{\mathrm{a,fwd}}$ & Min--max $E_{\mathrm{a,fwd}}$ & $\langle E_{\mathrm{a,rev}}\rangle$ & $\Delta E$ range \\
\hline
W$_8$B$_{10}$ & 4 & 7 & 7/0 & 1.22 $\pm$ 0.17 & 1.24 & 0.97--1.40 & 1.87 & $-0.98$ to $-0.21$ \\
W$_9$B$_9$ & 5 & 6 & 6/0 & 1.15 $\pm$ 0.25 & 1.17 & 0.74--1.41 & 1.63 & $-1.00$ to $+0.02$ \\
W$_9$B$_{20}$ & 6 & 11 & 11/0 & 1.37 $\pm$ 0.28 & 1.45 & 0.94--1.74 & 1.91 & $-0.95$ to $-0.07$ \\
W$_9$B$_{22}$ & 4 & 4 & 4/0 & 1.21 $\pm$ 0.29 & 1.16 & 0.93--1.60 & 1.86 & $-1.09$ to $-0.31$ \\
W$_{10}$B$_{21}$ & 3 & 3 & 3/0 & 1.32 $\pm$ 0.27 & 1.35 & 1.04--1.57 & 2.07 & $-1.04$ to $-0.53$ \\
W$_{14}$B$_{19}$ & 3 & 4 & 4/0 & 1.28 $\pm$ 0.27 & 1.33 & 0.90--1.54 & 1.53 & $-0.45$ to $0.00$ \\
W$_{16}$B$_{19}$ & 4 & 7 & 7/0 & 1.13 $\pm$ 0.33 & 1.18 & 0.70--1.53 & 1.43 & $-0.55$ to $+0.10$ \\
W$_{17}$B$_2$ & 5 & 7 & 4/3 & 1.08 $\pm$ 0.27 & 1.01 & 0.79--1.48 & 1.38 & $-0.55$ to $-0.01$ \\
W$_{17}$B$_{14}$ & 10 & 17 & 16/1 & 1.17 $\pm$ 0.25 & 1.24 & 0.65--1.57 & 1.71 & $-1.03$ to $-0.16$ \\
W$_{17}$B$_{17}$ & 9 & 20 & 20/0 & 1.37 $\pm$ 0.40 & 1.29 & 0.65--2.27 & 1.50 & $-0.91$ to $+0.60$ \\
W$_{18}$B$_9$ & 9 & 12 & 11/1 & 1.14 $\pm$ 0.35 & 1.24 & 0.40--1.54 & 1.53 & $-1.31$ to $+0.23$ \\
\end{tabular}
\end{ruledtabular}
\end{table*}

The representative energy profile for W$_{16}$B$_{19}$ in Fig.~\ref{fig:neb_barriers}a illustrates the general energetic sequence of the first N--H cleavage step. Molecular NH$_3^*$ adsorption is exothermic relative to the separated reactants, the climbing-image configuration marks the maximum along the calculated reaction path, and the dissociated NH$_2^*$ + H$^*$ final state is lower in energy than the corresponding molecularly adsorbed state. The barrier distributions in Fig.~\ref{fig:neb_barriers}b show that the first N--H bond cleavage is strongly pathway-dependent: individual calculated pathways for a given W--B composition span a broad range of both forward and reverse activation barriers. The widths of the boxplots and the scatter of the individual data points demonstrate that the reaction energetics are governed not only by the overall cluster composition, but also by the specific NH$_3$ adsorption geometry and the availability of a suitable nearby adsorption site for H$^*$.

Detailed reaction pathway analysis is focused on clusters with at least three sampled NH$_3$ initial states.
This threshold is introduced to ensure that the selected compositions provide enough local W--B environments to evaluate how site geometry affects the first N--H bond cleavage step. Table~\ref{tab:neb} summarizes the sampling statistics and barrier distributions for these compositions, and the corresponding multichannel reaction-energy profiles are shown in Fig.~S2 of the Supporting Information. The composition-level statistics depend on the sampled pathways, and the ranking of compositions by mean barrier is not preserved when the median or minimum values are considered; we therefore do not use the mean values to rank the intrinsic activity of individual compositions. For example, W$_{17}$B$_{17}$ shows one of the highest mean barriers but also contains pathways with barriers of approximately 0.65~eV. The relatively large standard deviations, especially for W$_{17}$B$_{17}$, W$_{18}$B$_9$, and W$_{16}$B$_{19}$, indicate that the barrier is not determined only by composition. Several pathways starting from the same NH$_3$ adsorption site lead to different final NH$_2^*$ + H$^*$ configurations and different activation barriers. 
The broad range of electronic activation barriers shows that the first N--H cleavage step depends strongly on the local arrangement of adsorption and H-accepting sites. Molecular NH$_3$ generally binds more strongly at W-top sites, whereas B is the nearest H-accepting atom in 93 of the 98 pathways summarized in Table~\ref{tab:neb}. These observations suggest that neighbouring W and B sites can play complementary roles during initial NH$_3$ activation. The calculations therefore identify local W--B coordination, rather than cluster composition alone, as a useful variable for selecting structures for further study. The reported barriers are electronic-energy differences; vibrational corrections have not been evaluated, so close barrier rankings and quantitative rate predictions remain unresolved.

The reverse barriers are generally larger than the forward barriers, typically reaching approximately 1.4--2.1~eV. This indicates that the dissociated NH$_2^{*}$ + H$^{*}$ state is generally stabilized relative to molecularly adsorbed NH$_3^{*}$, making recombination less favourable.

For context, the first N--H cleavage barrier was reported to be 0.85~eV for Ru$_{13}$/CNT \cite{zhou2018first}, approximately 1.25~eV for Pt$_3$/graphene \cite{cui2019enhanced}, 1.29~eV for bare Ni$_6$ \cite{miao2021graphene}, and 1.48~eV for Fe$_{55}$ \cite{lanzani2010nh3}. The W--B nanoclusters studied here show barriers in the same general range. However, these values should be compared qualitatively, since the reported systems differ in cluster size, support, exchange--correlation functional, reference structures, pathway-selection procedure, and treatment of the reaction energetics. In particular, the Pt$_3$/graphene value corresponds to a free-energy barrier including zero-point vibrational energy and entropy corrections, whereas the W--B barriers reported here are electronic-energy differences obtained from CI-NEB calculations.

The present analysis is limited to the initial activation stage of ammonia decomposition, namely molecular NH$_3$ adsorption and the first N--H bond-cleavage step. The overall catalytic performance may also depend on the energetics and kinetics of subsequent NH$_x$ dehydrogenation steps, migration of N-containing intermediates, N--N recombination, and H$_2$/N$_2$ formation and desorption. Accordingly, the barriers reported here should be interpreted as descriptors of initial NH$_3$ activation rather than as a complete assessment of ammonia-decomposition activity.

\section{Conclusions}

In this work, we combined evolutionary structure searches with density functional theory to investigate low-energy W$_m$B$_n$ nanoclusters and the initial stages of ammonia activation. The calculated $\Delta_{\min}$ and $E_{\mathrm{frag}}$ descriptors reveal a strongly non-monotonic stability landscape, with several compositions showing enhanced local stability relative to neighbouring stoichiometries. These descriptors characterize relative energetic trends within the explored isolated-cluster model.

NH$_3$ and H adsorption show pronounced composition- and site-dependent behaviour. NH$_3$ binds more strongly at W-top sites overall, although adsorption at B-top sites is also observed. The broad variation in adsorption energies within and between compositions demonstrates that adsorption cannot be described by the identity of the host element alone and is strongly influenced by the local atomic environment. A similarly broad site dependence is found for H adsorption, highlighting the structural diversity of the W--B cluster surfaces.

CI-NEB calculations further reveal a strong pathway dependence of the first N--H bond-cleavage step. For the selected multichannel compositions, the mean forward barriers are approximately $1.1$--$1.4$~eV, while substantial differences occur between individual pathways. The reaction energetics depend on both the initial NH$_3$ adsorption geometry and the local environment available to accommodate the detached H atom, emphasizing the importance of local structure in determining the energetics of NH$_3$ activation.

The present study therefore characterizes the initial activation of NH$_3$ on isolated W--B nanoclusters rather than the complete ammonia-decomposition reaction network. Subsequent NH$_x$ dehydrogenation, formation and desorption of H$_2$ and N$_2$, and the response of the clusters to reactive environments may further influence the overall catalytic behaviour. Within this scope, the results identify W--B nanoclusters as compositionally and structurally tunable candidates for further investigation of ammonia activation and provide an atomistic basis for extending the analysis toward the complete reaction mechanism.

\section{Data availability}

The data supporting the findings of this work are available in the GitHub repository:
\url{https://github.com/AlexanderKvashnin/WB_clusters}.
The repository contains the USPEX structure-search data, tabulated NH$_3$ and H adsorption-energy data, CI-NEB reaction-pathway data, representative NEB trajectories in XYZ format, and VASP input-setting information used for the analyses presented in the main text and in the Supporting Information.

\section{Funding Declaration}
This research received no specific grant from any funding agency in the public, commercial, or not-for-profit sectors. 


\bibliography{references}

\end{document}